\documentclass[11pt]{article}

\usepackage{amsmath}
\usepackage{amssymb}
\usepackage{mathtools}
\usepackage{amsthm}
\usepackage{graphicx}
\usepackage{multicol}

\usepackage[english]{babel}
\usepackage{fancyhdr}
\usepackage{color}
\usepackage{bbm}
\usepackage{natbib}
\usepackage{wrapfig}
\usepackage{algorithmicx}
\usepackage{algorithm}
\usepackage{booktabs}
\usepackage[noend]{algpseudocode}
\usepackage[font={small,it}]{caption}
\usepackage{float}
\usepackage[section]{placeins}
\usepackage{sidecap}
\usepackage{tikz}
\usetikzlibrary{decorations.markings}
\usepackage{comment}
\usepackage{soul}

\usepackage{hyperref}
\usepackage[margin=1in]{geometry}
\hypersetup{colorlinks,citecolor=blue,urlcolor=blue,linkcolor=blue}

\usepackage{listings}
\usepackage{color}

\definecolor{dkgreen}{rgb}{0,0.6,0}
\definecolor{gray}{rgb}{0.5,0.5,0.5}
\definecolor{mauve}{rgb}{0.58,0,0.82}

\makeatletter
\def\BState{\State\hskip-\ALG@thistlm}
\makeatother

\newtheorem{theorem}{Theorem}

\newtheorem*{lemma*}{Lemma}

\newtheorem*{remark}{Remark}

\theoremstyle{definition}

\newcommand{\e}{\varepsilon}

\newcommand{\p}{\mathbb{P}}

\DeclareMathOperator{\var}{Var}

\DeclareMathOperator{\cov}{Cov}
\DeclareMathOperator{\vecdiv}{Div}

\begin{document}

\title{Representation-Aware Experimentation: \\
Group Inequality Analysis for A/B Testing and Alerting} 
\author{Rina Friedberg \thanks{The authors would like to thank Shan Ba, Patrick Driscoll, Kenneth Tay, and YinYin Yu for thoughtful reviews and feedback on the manuscript. We are grateful for support from Parvez Ahammad and Meg Garlinghouse, and useful comments from Kinjal Basu, Karthik Rajkumar, and Wentao Su. Min Liu, Xiaofeng Wang, and Simon Yu contributed crucial design and engineering efforts to build the causal tree tool at LinkedIn. Guillaume Saint-Jacques is currently at Apple, Inc., and all work was done while he was at LinkedIn Corporation.} \\ \small{LinkedIn Corporation} \and Stuart Ambler \\ \small{LinkedIn Corporation} \and Guillaume Saint-Jacques \\ \small{LinkedIn Corporation}}
\date{\today}
\maketitle

\abstract{
As companies adopt increasingly experimentation-driven cultures, it is crucial to develop methods for understanding any potential unintended consequences of those experiments. We might have specific questions about those consequences (did a change increase or decrease gender representation equality among content creators?); we might also wonder whether if we have not yet considered the right question (that is, we don't know what we don't know). Hence we address the problem of unintended consequences in experimentation from two perspectives: namely, pre-specified vs. data-driven selection, of dimensions of interest. For a specified dimension, we introduce a statistic to measure deviation from equal representation (DER statistic), give its asymptotic distribution, and evaluate finite-sample performance. We explain how to use this statistic to search across large-scale experimentation systems to alert us to any extreme unintended consequences on group representation. We complement this methodology by discussing a search for heterogeneous treatment effects along a set of dimensions with causal trees, modified slightly for practicalities in our ecosystem, and used here as a way to dive deeper into experiments flagged by the DER statistic alerts. We introduce a method for simulating data that closely mimics observed data at LinkedIn, and evaluate the performance of DER statistics in simulations. Last, we give a case study from LinkedIn, and show how these methodologies empowered us to discover surprising and important insights about group representation. Code for replication is available in an appendix.
}

\section{Introduction}

A/B testing is a powerful tool. It allows us to understand whether changing a ranking algorithm improves search results,  sending a notification increases app visits, or a bug fix that should do nothing has any surprising results. When possible, A/B testing is the gold standard for understanding treatment effects, and therefore companies are increasingly using A/B tests to make decisions \citep*{10.1145/3292500.3330769, 10.1145/3377813.3381349, 10.1145/3447548.3467091}. LinkedIn currently runs hundreds of A/B tests per day, analyzing the impact of both major and minor adjustments, and always attempting to make data-driven decisions built on insights from experiments \citep*{10.1145/2783258.2788602}.

Given how central A/B testing is to decision making, for all of these experiments, a crucial follow-up question concerns their impact on different groups \citep*{DBLP:journals/corr/abs-2002-05819}. Sometimes experiments specifically aim to help underprivileged and under-represented populations succeed on LinkedIn; others have unrelated goals, but could still have unintended effects on representation. Moreover, factors such as human agency, bias in society, different preferences, etc., can all inform how different groups behave on LinkedIn and respond to changes on the platform. We can think about representation from many perspectives: demographic groups, members with different types and sizes of networks, employment history, etc. 
With so many experiments running simultaneously, a practical solution must automatically identify experiments where these differences emerge.

This question introduces a set of major statistical and practical challenges. First, how do we quantify consequences on representation in the context of a professional social network? Quantifying fairness and equity in a practically meaningful way has been a major challenge in similar research areas \citep*{CorbettDavies2018TheMA}, and hence defining a statistic that captures our quantity of interest is a key focus of this work. We approach this problem by answering the following key question: does an intervention bring a group closer to or further away from parity? As a toy example, if women represent 80\% of the LinkedIn member base, under parity, they would also represent 80\% of confirmed hires. If women in fact represent 50\% of confirmed hires, then any experiment that increases that percentage - even if that experiment benefits men more than women - brings us closer to parity. 
Even in this illustrative example, we can consider the complicated dynamics at play; there could be many reasons outside of the LinkedIn platform that could cause a difference in rates of confirmed hires.

Observe that demographic parity is not necessarily the same thing as fairness, or freedom from bias; different groups may have different preferences, which can then be revealed in product metrics. 
Furthermore, fairness and representation are not the same concept; indeed, there are many ways to define fairness-related goals for a given system, almost always incompatible with one another \citep*{kleinberg2017inherent}. 
However, representation is useful for us as a first approximation to call for a deeper look into certain experiments. 
We formalize this intuition with deviation from equal representation (DER) statistics in Section \ref{sec:derstats}. 

Second, how should we consider other dimensions of interest for highly impactful experiments? DER statistics can tell us whether an A/B test increased representation for a specific set of groups (for example, genders); but we need to specify the groups before we can calculate the statistics. We may not always anticipate which features from a known set determine which groups have a different experience under a given treatment, or how those features might interact. Therefore, we complement the DER statistics with a search for heterogeneous treatment effects that does not require a specific hypothesis. The most useful results for applications at LinkedIn would be cohort-level results: we want to know if there were groups of members, advertisers, etc. that had a different experience from other groups. An alternative to this would be estimating individual-level treatment effects (see for example \citet*{lassohte, causalforest, bart}). Our priority is results that are straightforward to interpret and act on, and we therefore suggest a causal tree \citep*{causaltree}, which searches along given features to return a set of cohorts with heterogeneous treatment effects. 

Third, how do these solutions fit into a large-scale experimentation platform? A/B testing at scale introduces many interesting challenges \citep*{Kohavi:2013:OCE:2487575.2488217, 10.1145/2566486.2567967, NEURIPS2020_1e0b802d}, and useful results need to account for the structure of our experimentation system and the broader goals. At LinkedIn, the scale of A/B tests we run and metrics we calculate makes searching for heterogeneous treatment effects along every possible experiment, treatment variant, metric, etc. unrealistic. DER statistics are far faster to compute, but in identifying experiments with an extreme impact on representation we need to consider practical significance and multiple testing effects. Avoiding false discoveries is particularly important, as we need to develop a trustworthy system in order to drive impact internally. 

At LinkedIn, these solutions were developed and implemented in a platform called Every Member, with the goal of supporting LinkedIn's mission to bring economic opportunity to \textit{every member} of the global workforce. Every Member has the ability to calculate DER statistics on different groups for a large number of experiments and a large set of important metrics, such as job applications, invitations sent, messages received, etc., in our experimentation system. 
It sends alerts for experiments with extreme impact. These experiments are sometimes directly related to equity goals, but often are unrelated, with completely unintended consequences. Learning about these consequences is crucial for Every Member's ability to serve its purpose.

\citet*{pmlr-v81-buolamwini18a}, in a cornerstone paper on algorithmic fairness, emphasize the importance of accountability and proactive auditing in major systems. Their focus is on facial recognition systems and computer vision, but their message applies directly to A/B testing; proactive monitoring for impact on representation must be a key priority for any experimentation system. Similarly, \citet*{10.1145/2939672.2945386} emphasize the importance of fairness-aware data mining in algorithmic bias prevention and mitigation; these principles again apply to experimentation best practices.

There is a fast-growing literature describing fairness in large-scale data analysis, particularly in regards to algorithmic bias. Previous work on fairness in classification models has introduced methods such as utility-maximization subject to fairness constraints \citep*{10.1145/2090236.2090255}, permutation tests to flexibly evaluate bias \citep*{10.1145/3394486.3403199}, and fairness metrics to account specifically for distributional differences between groups \citep*{beutel}.
\citet*{10.1145/3351095.3372878} consider long-term fairness impact of machine learning decision systems and introduce simulation frameworks to give a more complete picture. More closely related to our work, \citet*{10.1145/3219819.3220046} study group effects, decomposing algorithmic unfairness into between-group and within-group effects, and noting tradeoffs therein.

The remainder of our paper is structured as follows. Section \ref{sec:methods} introduces the DER statistic and its Central Limit Theorem, and discusses how we use causal trees for this application. Section \ref{sec:implementation} details how we run simulations to evaluate the DER alerting method and gives simulation results. Section \ref{sec:casestudy} gives a case study from LinkedIn, describing a broad initiative that was flagged by the DER alerted method and discussing heterogeneous treatment effects we discovered. Section \ref{sec:conclusion} concludes.

\section{Methodology}\label{sec:methods}

\subsection{Pre-Specified Dimensions: DER Statistic}\label{sec:derstats}

In this section, we introduce the DER statistic to measure distance from a form of demographic parity for $k$ pre-specified groups, for example genders, by comparing the metric means. Suppose each group has mean $\mu_i$, for $i \in \{1, \dots, k\}$. Define the population-level DER statistic $D$ as
\begin{align}
D(\mu_1, \dots, \mu_k) &= \frac{k}{k-1} \sum_{i=1}^k \left( \frac{\mu_i}{\sum_{j=1}^k \mu_j} - \frac1k \right)^2 
\end{align}
Note that here we assume the metric is non-negative.

This statistic has several key features to emphasize. First, we can define one statistic that simultaneously compares all $k$ groups, as opposed to running $k(k-1)/2$ comparisons. Second, we do not need to choose a particular under-represented group. Third, the value is relative to the overall level of the underlying metric, as opposed to absolute, which means we can naturally compare results across metrics of different scales. Fourth, note that the metric is squared, to emphasize shares far from parity. 

We can also bookend this metric by considering two cases. First, suppose all metric means are equal. Then, $\mu_i/\sum_{j=1}^k \mu_j = 1/k$ for all $i$, and the statistic is exactly 0. On the other extreme, suppose all $\mu_i = 0$, except for $\mu_1$. In this case, 
\begin{align*}
D(\mu_1, \dots, \mu_k) &= \frac{k}{k-1} \left( \left(1 - \frac1k\right)^2 + (k-1) * \left(\frac1k\right)^2 \right) \\
&= \frac{k-1}{k} + \frac1k = 1.
\end{align*}
That is, $D(\mu_1, \dots, \mu_k)$ is exactly zero when all group means are equal, and is exactly one when only one group has a non-zero mean.

We will estimate $D(\mu_1, \dots, \mu_k)$ using plug-in estimators for group means, so that $\hat{D} = D(\widehat{m_1}, \dots, \widehat{m_k}),$ where $\widehat{m_j}$ is the observed sample mean. 
Now suppose we have an A/B test with two variants, control (variant 0) and treatment (variant 1), and assume that each group $k$ has at least one member assigned to each variant. We are interested in comparing $D(\mu_1^1, \dots, \mu_k^1) - D(\mu_1^0, \dots, \mu_k^0)$, where superscript denotes assignment to treatment or control. In practice, we will observe group averages $\widehat{m_i}$, and alert based on the natural statistic $\widehat{\Delta}$ formed by substituting $\widehat{m_i}$ as an estimate for $\mu_i$.
\begin{align}
\widehat{\Delta} &= \frac{k}{k-1} \sum_{i=1}^k \left( \frac{\widehat{m_i}^1}{\sum_{j=1}^k \widehat{m_j}^1} - \frac1k \right)^2 - \frac{k}{k-1} \sum_{i=1}^k \left( \frac{\widehat{m_i}^0}{\sum_{j=1}^k \widehat{m_j}^0} - \frac1k \right)^2 
\end{align}

\begin{figure}[t]
\centering
\includegraphics[width=0.6\textwidth]{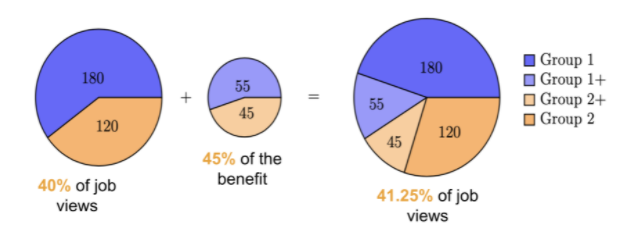}
\caption{Visualization of an intervention that brings group 1 and group 2 closer to parity, even though the intervention itself favors the over-represented group.}\label{pie-image}
\end{figure}

We noted earlier that one of our goals with this metric is to identify experiments that bring us closer to parity, even if they benefit an over-represented group more than an under-represented group. We briefly focus on this point, giving a visual explanation in Figure \ref{pie-image} using a toy example. Suppose we have two groups (purple and orange) of equal numbers of people, where at baseline group 2 contributes 40\% of daily job views on LinkedIn. Consider an intervention, shown as the small central image, in which the members of group 2 who received the intervention contributed 45\% of daily job views. Note that group 2 actually does worse than group 1 under the intervention (45\% of the benefit, as opposed to 55\%); but because on average, group 2 has increased job views under the intervention compared to the baseline, the change overall brings the two groups closer to parity, as shown in the final image. The DER statistic decreases from $0.04$ to $0.03$ due to the intervention.

In order to use DER statistics in practice, we need to quantify uncertainty and build confidence intervals. We now move to give a Central Limit Theorem for DER statistics and an empirical validation for their finite sample performance.

\subsubsection{Asymptotic Distribution}

Theorem \ref{th-clt} establishes that the DER statistic $D(\widehat{m}_1, \widehat{m}_2)$ has an asymptotically Normal distribution and gives a variance estimate, both derived from the Delta Method. We give the theorem for $k=2$ groups for simplicity; results for $k\ge 2$ groups can be derived analogously.

\begin{theorem}[Central Limit Theorem]\label{th-clt} Suppose we have a sample of n individuals drawn from a hypothetical infinite population, where each individual $i$ belongs to a group $G_i \in \{1, 2\}$ and has a response $Y_i \in \mathbb{R}$. For each group $g$, denote the sample conditional mean by $\widehat{\mu_g}$, population conditional mean by $\mu_g$, population conditional variance by $\sigma_g^2$, and population fraction by $p_g$ (so that $p_1 + p_2 = 1$). Then, 
\begin{align}
&\sqrt{n}(D(\widehat{\mu}_1, \widehat{\mu}_2) - D(\mu_1, \mu_2) ) \to_D \mathcal{N}(0, \sigma^2), \nonumber
\intertext{where the asymptotic variance $\sigma^2$ is}
\sigma^2 &= \frac{16 (\mu_1 - \mu_2)^2}{(\mu_1 + \mu_2)^6} \left( \mu_2^2 \left( \frac{\sigma_1^2}{p_1} \right) + \mu_1^2 \left( \frac{\sigma_2^2}{p_2} \right) \right) \label{dm_der_variance}
\end{align}
\end{theorem}

A proof can be found in Appendix \ref{app-proof}. 
Observe that the covariance matrix $\Lambda^2$, defined for $0 < p_1, p_2 < 1$, is singular when and only when its determinant is zero, which occurs if and only if at least one of $\sigma_1^2$ and $\sigma_2^2$ is zero (the metric is constant for at least one $g$). Except in that unusual case, the Delta method variance is zero only in case of equality, $\mu_1 = \mu_2$.  In practice, exact equality is unusual. Observe also that our proof fixes $k=2$; an analogous result for more general $k$ can also be obtained, but is omitted for simplicity. 

\subsubsection{Finite Sample Performance}

We will use the results of Theorem \ref{th-clt} to build Gaussian confidence intervals for statistics $\widehat{\Delta}$ capturing the change in DER between variants of an A/B test. While these will be asymptotically valid confidence intervals, we also want to validate that they perform reasonably well on finite samples. 
To do this, we will compare variance estimates from (\ref{dm_der_variance}) with estimates from empirical resampling methods, namely the bootstrap and permutations. These methods might be the first choice for variance estimation, but they will not scale easily given how many statistics we need to evaluate; so we use them as a comparison benchmark in simulations.

\begin{figure}[t]
\centering
\includegraphics[width=0.5\textwidth]{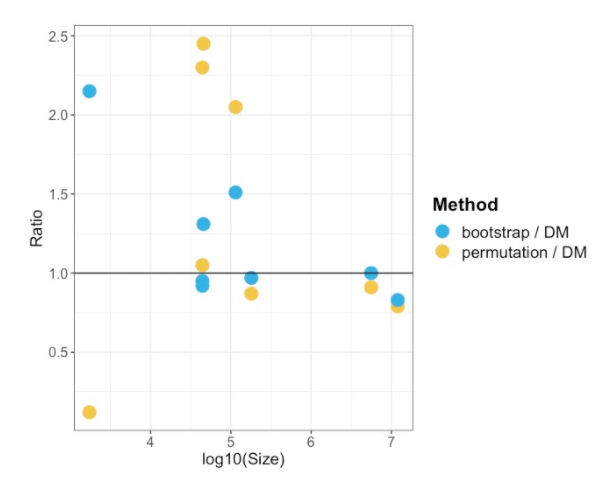}
\caption{Ratio of bootstrap and permutation variance estimates to Delta Method on a set of LinkedIn experiments, by size of experiment population.}\label{var-image}
\end{figure}

 We take a set of previously identified, highly impactful experiments, and compare the asymptotic variance estimate from Theorem \ref{th-clt} with resampling variance estimates (Figure \ref{var-image}). On experiments of 40,000 members (relatively small), the estimates were off by a maximum factor of 2.5. As the sample size increases, observe that the delta method variance estimates approach the bootstrap and permutation variance estimates; in other words, the asymptotic result kicks in. 

\subsection{Search for Relevant Groups: Causal Tree}\label{sec:causaltree}

A complementary methodology aims to answer the question: are there any dimensions or combinations thereof such that, if we were to group members along these dimensions, we would see a large difference in responses between the treatment arms? Instead of choosing a specific group delineation such as gender, we search among a wide set of possible features to discover relevant group definitions in a data-driven way. 

Here we briefly describe our implementation of the causal tree algorithm \citep*{causaltree}. The algorithm recursively partitions the observed units in a training dataset into two groups, dividing within those groups again, until the dataset has been split into many small cohorts. The authors give several versions of this algorithm; we use the squared t-statistic criterion, which searches for the greatest evidence of different treatment effects between two groups resulting from a new split. To prune the tree, we choose the largest partition such that the confidence intervals for group-wise average treatment effects are non-overlapping. We then re-estimate all cohort-level treatment effects and corresponding confidence intervals on an estimation dataset. 

Our target here is relative lift; that is, for a metric with control group average $\bar{Y}^0$ and treatment group average $\bar{Y}^1$, we are interested in $\widehat{\xi} = (\bar{Y}^1 - \bar{Y}^0)/\bar{Y}^0$. To build appropriate confidence intervals here, observe that the sample means for the groups follow an asymptotically Gaussian distribution. 
\begin{equation*}
\begin{pmatrix} \bar{Y}^0 \\ \bar{Y}^1 \end{pmatrix} - \begin{pmatrix} \mu_0 \\ \mu_1 \end{pmatrix} \to_D \mathcal{N} \left(\begin{pmatrix} 0 \\ 0 \end{pmatrix}, \begin{pmatrix} \sigma_0^2/n_0 & 0 \\ 0 & \sigma_1^2/n_1 \end{pmatrix} \right). 
\end{equation*}

We can then again use the Delta Method to establish that our relative lift estimates are asymptotically Gaussian, with variance
\begin{align*}
\sigma_r^2 &= \frac{\mu_1^2}{\mu_0^4} \frac{\sigma_0^2}{n_0} + \frac{1}{\mu_0^2} \frac{\sigma_1^2}{n_1}
\intertext{Hence to build confidence intervals, plug in the sample averages $\bar{Y}^w$ and standard deviations $s_w^2$, for $w \in {0,1}$.}
\widehat{\sigma_r}^2 &= \frac{\bar{Y}_1^2}{\bar{Y}_0^4} \frac{s_0^2}{n_0} + \frac{1}{\bar{Y}_0^2} \frac{s_1^2}{n_1}
\end{align*}

In practice, this pruning procedure often yields an empty tree; that is, there is no division suggested by the causal tree in which the cohorts have significantly different relative lift estimates. This is not a bug; often heterogeneous treatment effects are very difficult to detect, and if we are going to use results moving forward, we should be confident that the groups we detect are indeed having different experiences. In other use cases, one might want this method to be more exploratory, in which case we would suggest a less stringent pruning criterion.

\section{Implementation in a Large-Scale Experimentation System}\label{sec:implementation}

To put this methodology into practice, we first need to consider the ecosystem in which our experiments run. Many experiments run simultaneously, comparing several variants (often more than just treatment vs. control) and evaluating the variants against many benchmark metrics. Hence we need to evaluate a large set of comparisons with complex correlation structure. For each experiment, pair of variants, and metric, we calculate the difference of DER statistics between the two variants. 

In using those results to alert about high-impact A/B tests, we use a Bonferroni Correction to identify highly significant results. Suppose we have n total comparisons, including multiple comparisons from the same experiment and treatment variant. We select as significant alerts all combinations of experiment $i$, metric $j$, and treatment variants $\{k, \ell\}$ such that the corresponding p-value, $p_{i,j,k,\ell}$, satisfies 
\[p_{i,j,k,\ell} \le \alpha/n,\]
with the significance value $\alpha$ set as 0.05. 

The key appealing feature of the Bonferroni correction is that it protects against family-wise error rates (that is, the probability of making at least 1 false discovery) under any correlation structure \citep*{10.1093/aje/kwy250}. We need high confidence that the experiments we identify have statistically significant impact, in order to build trust with experiment owners and keep our guardrails highly reliable. A natural concern, however, is that if we use such a stringent correction, we need to make sure we have sufficient power to detect experiment effects in context. The next section therefore addresses building simulations that give believable guarantees on the family-wise error rate and power of the DER alerting system. 

\subsection{Simulating LinkedIn Data}

When building our simulations, we have the following set of goals: (i) evaluate false discovery rate and power of the DER alerting procedure as it would work for a given week of experiments at LinkedIn; (ii) build a replicable simulation system, useful both for validation within LinkedIn and for experimentation for external practitioners. Building such a simulation setup involves designing a data-generating process that can generate a dataset mimicking the results of LinkedIn experiments. 
We observe that most results (for example, member sessions last week) will have a spike at zero, a distribution around smaller numbers, and a long right tail. This suggests that we are looking for mixture distributions; a point mass at zero, and a long-tailed distribution. 

To identify the best mixture distributions, we compare several candidate distributions (Weibull, Log-Normal, and Levy), with mixtures of a point mass at zero. We also draw vectors of responses to past LinkedIn experiments. We then perform a grid search over parameters for the mixture distributions, and the distributions themselves, comparing the generated data to the observed data using a Kolmogorov-Smirnov test \citep*{10.2307/2280095}. Over many different vectors of data, we discovered a best-fit distribution -- the Log-Normal mixed with a point mass at zero -- and an appropriate range of parameters, that generate samples approximating LinkedIn data. 

This builds our simulation procedure. For each of $N$ A/B tests to simulate, we simulate $m$ metrics by drawing hyper-parameters from the given range, and draw a Log-Normal mixed with a point mass at zero, according to those values. We can add in treatment effects with the desired magnitude. 

{\bf Step 1.} Draw parameters.
\begin{align*}
\text{Data Size}~ n &\sim \mathcal{U}(a_1, a_2) \\
\text{Overall mean}~ \mu &\sim \mathcal{N}(b_1, b_2) \\
\text{Group effect}~ \alpha &\sim \mathcal{U}(c_1, c_2) \\
\text{Treatment effect}~ \beta &\sim \mathcal{N}(d_1, d_2) \\
\text{Group/treatment interaction} ~\delta &\sim \mathcal{N}(e_1, e_2) * \text{Bernoulli}(\e) \\
\text{Mixture probability} ~p_j &\sim \mathcal{U}(f_1, f_2)\\
\text{Standard deviation}~ \sigma &\sim \mathcal{U}(g_1, g_2)
\end{align*}

We also randomly assign group and treatment to each individual in the sample, and fix the fraction $\e$ of non-zero group/treatment interaction effects, where $W$ denotes the treatment. 

{\bf Step 2.} Draw Log-Normal mixture responses as follows.
\begin{align*}
Y &\sim \begin{cases} \text{Log Normal}(\mu + \alpha * 1\{G =1\} ~+ \\
~~~~~~~~~ \beta * 1\{W = 1\} ~+  \\
~~~~~~~~~ \delta * 1\{G =1\} * 1\{W = 1\}, \sigma) & \text{~wp~} p \\ 0 & \text{~wp~} 1 - p \end{cases}
\end{align*}

Table \ref{tab-power} gives results from repeating this simulation procedure 100 times. 
We compare results from alerting based on DER statistics with alerting based on differences-in-differences analysis comparing treatment effects for two groups. 
Observe that non-zero DER statistic changes will correspond to non-zero group/treatment interaction effects, but depending on other parameters, large effects for one category may not correspond to large effects in the other (recall Figure \ref{pie-image}). 
We therefore quantify power in several ways; we report power against group/treatment interactions in general, and also on large effects in each category. 
For group/treatment interaction, this corresponds to $1\%$ of the control metric mean, averaged over groups. 
We also define $|D^*|$, the DER statistic computed on the expected values of the means, and report on the performance against representation changes by giving the rejection rate in the top quantiles of $|D^*|$. 

\begin{table}[t]
    \centering
    \begin{tabular}{lcccc}
     && DER statistics && Difference in group lifts \\
    \cline{3-3} \cline{5-5}
    $\%$ Rejections && 3.9\% && 24.2\% \\
    False discoveries on group/treatment interactions && 0\% && 80.9\% \\
    Power against group/treatment interactions && 73.9\%  && 78.3\% \\
    Power against group/treatment interactions $\ge 1\%$ && 85\% && 82.5\% \\
    Power against top $2.5\%$ of $|D^*|$ && 96\% && 100\% \\
    \end{tabular}
    \caption{Results from simulations evaluating full DER alerting methodology. Here we suppose we have 1,000 A/B tests, with a group/treatment interaction effect added with $5\%$ probability. We compare results from both group-level difference-in-differences (DID) and DER statistic searches, both with a Bonferroni correction and $alpha=0.05$, reporting false discoveries and power on group/treatment interaction terms, and rejection rates at large values of relative group/treatment effects and of $|D^*|$ (the DER statistic computed on the expected values of the sample means). The $97.5\%$ quantile for DER statistics (using expected values) is on average $0.0028$. Results are averaged over 100 runs.}
    \label{tab-power}
\end{table}

Using DER statistics under a Bonferroni correction, we reject over $80\%$ of large values of $|D^*|$ in simulated experiments, and do not make any false rejections. 
Difference-in-differences analysis, however, is essentially unusable due to extremely high false discovery rates, even having corrected for multiple testing, again with the Bonferroni correction. 
The phenomenon of difference-in-differences rejecting too many true null hypotheses (sometimes around 40\%), often in the case of time-varying data, has been established in the literature \citep*{10.1162/003355304772839588}. 
Bootstrap variance estimates can help mitigate this problem, but as previously discussed for DER statistics, are not a practical solution for us. 
These results are crucial for our confidence in the multiple testing correction, and the ability of the DER statistic to identify highly impactful experiments without major risk of false discoveries. 

\section{Case Study: Increasing Member Value}\label{sec:casestudy}

We now move to discuss a case study from the DER alerting system implemented at LinkedIn. 
There was a broad initiative at LinkedIn in 2021 to increase member value, as measured by increasing sessions, through focal areas such as encouraging content creation and supporting network growth. 
This initiative encompassed over 100 experiments over the course of months. 
To evaluate its overall impact, there was also a holdout experiment, in which 2\% of members at LinkedIn did not receive a treatment variant for any of the experiments. This was a crucial step to help us understand the overall impact of making so many changes under one umbrella. This holdout experiment was alerted as having a significant impact on gender representation\footnote{Gender representation is currently measured as men/women in the DER alerting system, due to data availability. Expanding the alerts to include other gender identities is an important step for future work.}.

\subsection{Pre-Specified Group: Gender Representation}

\begin{figure}[t]
    \centering
    \includegraphics[width=0.49\textwidth]{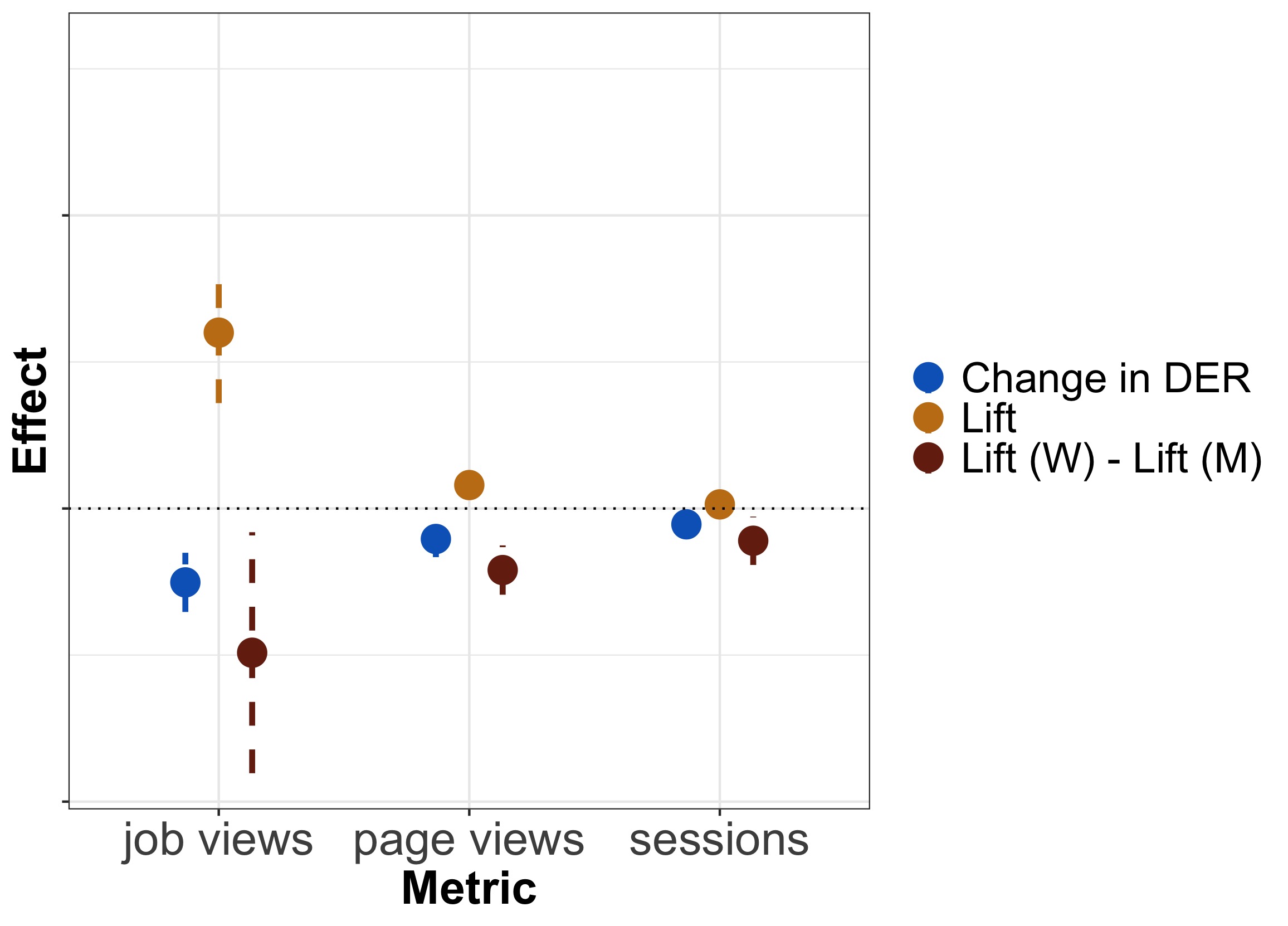}
    \includegraphics[width=0.49\textwidth]{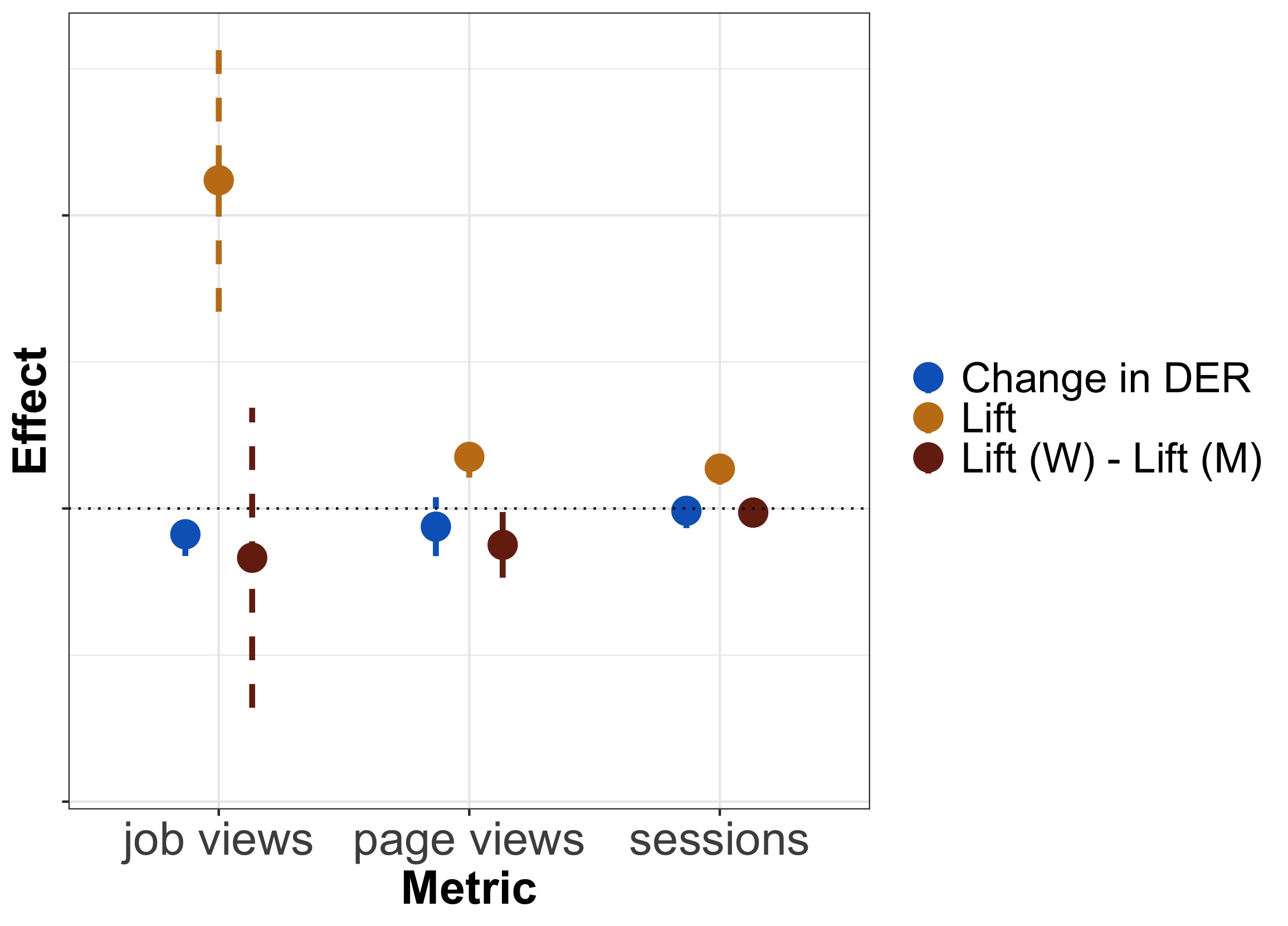}
    \caption{Gender representation impact of the holdout experiment on three key metrics, measured at two time periods. We show the overall lift, difference between (unsquared) DER statistics, $\widehat{\Delta}$, and difference between relative lift for women and men, $\widehat{\xi}_{\text{W}}$ - $\widehat{\xi}_{\text{W}}$. The first is between October 22, 2021 and November 7, 2021, corresponding to the DER alert; the second is between January 3, 2022 and January 31, 2022, corresponding to a stable period after all experiments have been ramped. The y-axis is standard across both figures; observe that the values for lift and for difference in lift for men and women are comparable, while the values for $\widehat{\Delta}$ should be compared only to each other. The thin dotted line denotes an effect of $0\%$.}
    \label{fig:der_alerts}
\end{figure}

Figure \ref{fig:der_alerts} gives results from the DER monitoring system on the overall holdout experiment. We show three outcomes; overall lift, the difference between DER statistics for treatment and control ($\widehat{\Delta}$), and the difference between relative lift for men ($\widehat{\xi}_M$), and for women ($\widehat{\xi}_F$). 
Note that the DER statistic as defined and as earmarked for alerts is squared; these results are the square root, for better interpretability. 
A negative value for $\widehat{\Delta}$ indicates that the experiment brought us further away from parity in gender representation; a negative value for $\widehat{\xi}_F - \widehat{\xi}_M$ indicates that the experiment benefited men more than women. 

Further, we show these values for two date ranges (left panel). 
The first date range, 10/22/21 through 11/7/21, corresponds to the alert that there was a significant impact on representation. We display results for three key metrics: job views, page views, and sessions. In these date ranges, we can observe significant and positive lift for job and page views, and negative impact on representation.

\begin{remark} 
Note that the patterns we observe here can be attributed to many potential causes, including unexpected differences between how different LinkedIn members behave, and existing differences in society; indeed, with this context, we would \textit{expect} to see some measured differences. Our tool helps us identify these differences and potentially target improvements in the future.
\end{remark}

When we move to the 1/3/22-1/31/22 date range (right panel), all experiments belonging to the member value initiative have been ramped, so we can interpret the results as being more stable. We can observe that perhaps due to additional changes implemented between November and January, lifts are more strongly positive overall for all three metrics, and the negative impact on representation has faded. The point estimates for group representation measures, however, remain negative. While new changes likely mitigated some of the representation gap increases, it is still crucial to understand where the discrepancy in gender experience is coming from. This brings us to the other component of our analysis: searching for cohorts of individuals who responded differently to the experiment, looking for new insights that can help us target future improvements.

\subsection{Search for Groups: Dormant Women Members}

\begin{figure}[t]
    \centering
    \includegraphics[width=0.49\textwidth]{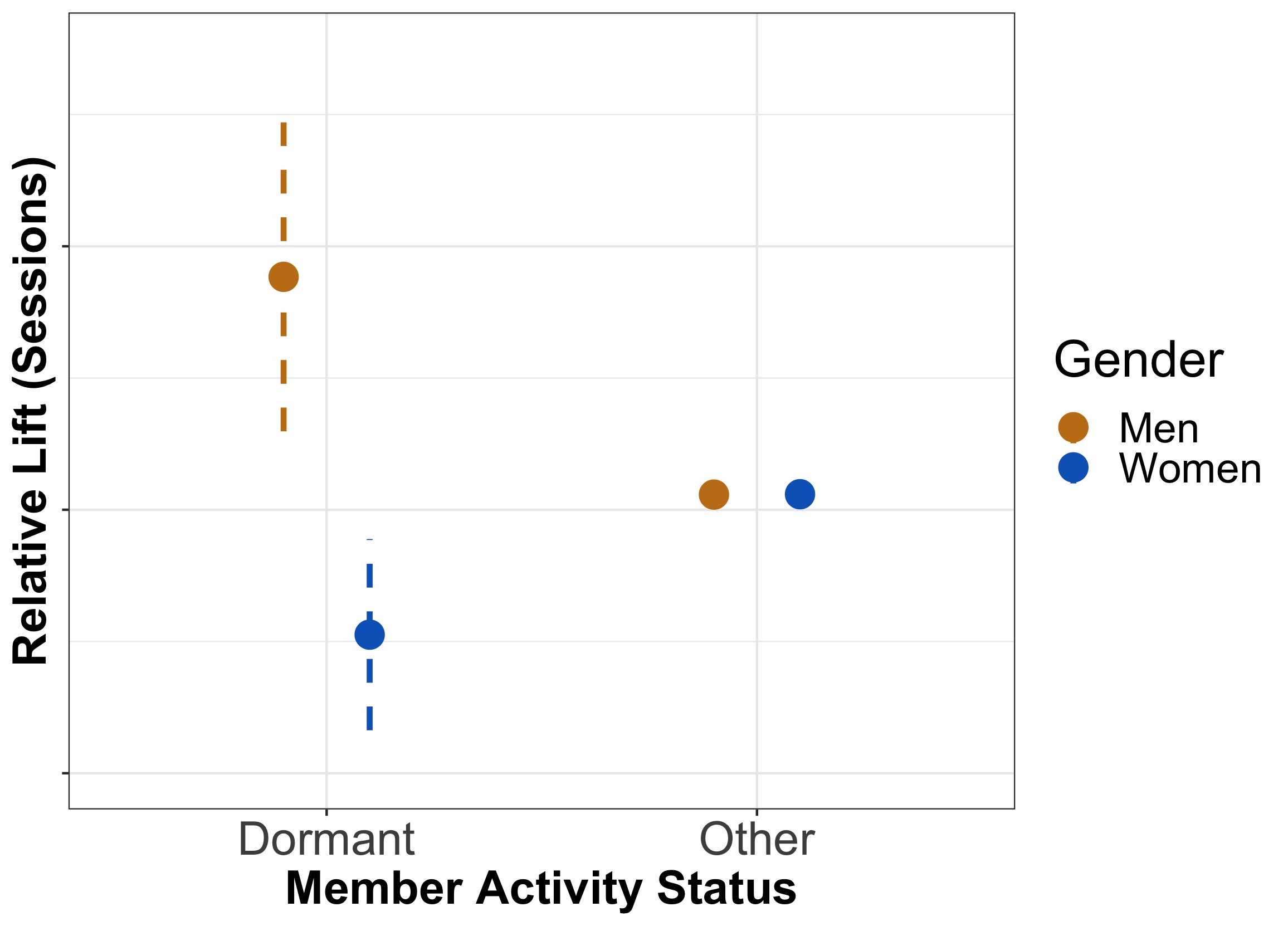}
    \includegraphics[width=0.49\textwidth]{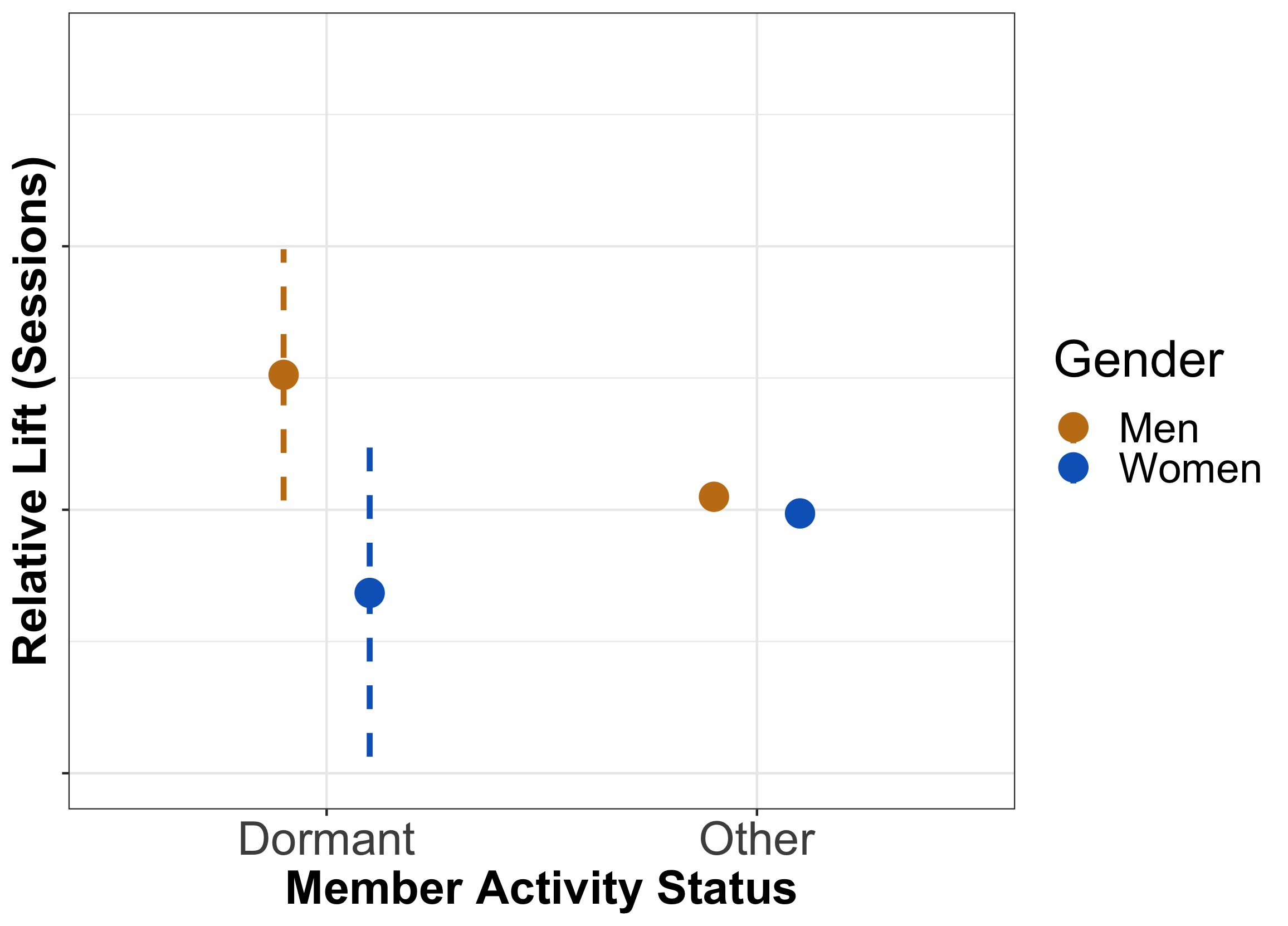}
    \caption{Relative lift on sessions for dormant members and for members with more recent activity, broken down by men/women. The left panel shows these heterogeneous treatment effects for the first time period, corresponding to the DER alert; the right panel, once all changes were added and stabilized in January. The y axis is standard across both figures.}
    \label{fig:causaltrees}
\end{figure}

To better understand the holdout experiment, we search for cohorts using the causal tree methodology described in Section \ref{sec:causaltree}. 
We consider the impact of the overall holdout on member sessions, as a proxy for member value, in the date range from 10/22/21 through 11/7/21. 
We considered a set of 15 features including member activity status, number of connections, country, and gender. 
Figure \ref{fig:causaltrees} shows the results from the causal tree in the left hand panel. 
We see that dormant women members had a relatively worse experience under the overall initiative, compared to dormant men members. 
Among more active members, men and women had similar relative lifts.  
This helps us develop a more detailed hypothesis for the results of the DER alerting system: dormant men members are responding more, while dormant women members actually visit less.

\begin{remark}
Similarly to our interpretation of DER statistics, there can be behavioral differences and other possible causes leading to the patterns in Figure \ref{fig:causaltrees}. Hence this result does not showcase unfairness in LinkedIn products directly, but highlights opportunities for future improvements.
\end{remark}

These patterns start to fade as we consider the relative lift for the same four cohorts in January, once all of the changes have been implemented and effects have had a chance to stabilize. We still see a discrepancy between men and women members who were dormant, but the confidence intervals for relative lift now overlap.

Viewed alongside the DER alerting results, this helps us to identify why those differences may have been seen, and why they changed over time. The patterns can come from many sources, but if we do not identify them, we cannot help. With these tools in hand, however, we can take a deep dive into the different behaviors and impacts we observed in this initiative, identifying which members may not be getting the full possible value, and improving value for them specifically in the future. 

\section{Discussion}\label{sec:conclusion}

In this paper, we introduced a method for monitoring large-scale experimentation systems for impact on group representation. We proposed DER statistics, which quantify whether A/B tests bring an arbitrary number of groups closer to or further from parity, and gave a Central Limit Theorem. We discussed practicalities of using DER statistics to alert about high-impact experiments, prioritizing user impact and trust, and provided a simulation scheme designed using prior data from metrics at LinkedIn. We complemented the DER alerting methodology with a slightly modified causal tree, which helps us to search for cohorts with heterogeneous treatment effects along unknown dimensions, as opposed to along pre-specified groups, directly comparing relative lift instead of DER changes. Last, we gave a case study from an initiative identified by the DER alerting system implemented at LinkedIn. 

As companies gather more data and run more experiments, responsible and equity-aware experimentation practices must be a cornerstone of data analysis and decision-making. 
Monitoring experiments for impact on group representation helps us to identify both features that decrease equal representation, and features that improve representation and can shed light on why. 
Moreover, we can understand how large-scale projects like our case study change over time, understanding the effect not only of individual experiments at one time, but of holdout projects and how they change over the months. 
It is worth noting that group representation is one goal for responsible experimentation; equity and fairness can be defined in a myriad of ways, and future work should consider different ways of quantifying these concepts as they apply to large-scale A/B testing and alerting. 
Ultimately, statistics and methodology like this are key for us as we guard against unintended consequences of A/B tests on our platform, leveraging statistical and computational tools to advocate for diversity of representation.

\bibliographystyle{apalike}
\bibliography{biblio}

\newpage

\appendix
\section{Proof of Theorem \ref{th-clt}}\label{app-proof}

\begin{proof} 
We will use the delta method to identify the asymptotic distribution of $\sqrt{n}(D(\widehat{\mu}_1, \widehat{\mu}_2)$. 
We proceed in two steps, first obtaining the asymptotic covariance matrix for the application of a standard Central Limit Theorem to the random variables $I_{G=g}$ and $I_{G=g} Y$, and second, finding the asymptotic distribution for $D(\widehat{\mu}_1, \widehat{\mu}_2)$ by applying the chain rule, using the derivative of a division operator giving the group conditional means from the expectations of those variables. 

We use the notation $f_X$ for the pdf of the random variable $X$, $I_A$ for the indicator variable for the event $A$, and $\p(A) = \mathbb{E}(I_A)$ for probability of $A$.  
Using standard definitions,
\begin{align}
  f_{I_{G=g} Y} = \p(G=g) f_{(Y | G=g)} = p_g  f_{(Y | G=g)} , \notag
\end{align}
where we set $p_g = \p(G=g)$. 

For each value $g$ of $G$ we obtain two corresponding random variables on the infinite population, $I_{G=g}$ and $I_{G=g} Y$. We also denote the other value of $G$ as $g^{\prime}$. Then, setting $m_g = \mathbb{E}(I_{G=g} Y)$ for later use,
\begin{align*}
  \mathbb{E}(I_{G=g}) &= p_g \\
  \var(I_{G=g}) &= p_g (1 - p_g) \\
  \mathbb{E}(I_{G=g} Y) &= m_g = p_g \mu_g \\
  \mathbb{E}((I_{G=g} Y)^2) &= p_g (\sigma^2_g + \mu_g^2) \\
  \var(I_{G=g} Y) &= p_g \sigma_g^2 + p_g (1 - p_g) \mu_g^2  \\
  \cov(I_{G=g}, I_{G=g} Y) &= p_g (1 - p_g) \mu_g \\
  \cov(I_{G=g}, I_{G=g^{\prime}}) &= - p_g p_{g^{\prime}} \\
  \cov(I_{G=g}, I_{G=g^{\prime}} Y) &= - p_g p_{g^{\prime}} \mu_{g^{\prime}}   \\
  \cov(I_{G=g} Y, I_{G=g^{\prime}} Y) &= - p_g p_{g^{\prime}} \mu_g \mu_{g^{\prime}} 
\end{align*}
From here, we build the population covariance matrix, composed of four $2\times 2$ block matrices. 

\begin{align*}
    \Sigma^2 &=  \cov(<I_{G=2}, I_{G=2}, I_{G=1} Y, I_{G=2} Y>) \\
    &= \begin{bmatrix} A & B \\
    B^T & C \\
    \end{bmatrix} 
    = \begin{bmatrix}
        p_1 (1 -p_1) & -p_1 p_2 &  p_1 (1 -p_1) \mu_1 & -p_1 p_2 \mu_2 \\
        -p_1 p_2 & p_2 (1-p_2) & -p_1 p_2 \mu_1 & p_2 (1-p_2) \mu_2 \\
        p_1 (1 -p_1) \mu_1 & -p_1 p_2 \mu_1 & p_1 \sigma_1^2 + p_1 (1-p_1) \mu_1^2 & -p_1 p_2 \mu_1 \mu_2 \\
        -p_1 p_2 \mu_2 & p_2 (1-p_2) \mu_2 & -p_1 p_2 \mu_1 \mu_2  & p_2 \sigma_2^2 + p_2 (1-p_2) \mu_2^2 \\ 
    \end{bmatrix}
\end{align*}

Now, we note the central limit theorem for the sample averages $\widehat{p_g}$ and $\widehat{m_g}$.
\begin{align*}
  \sqrt{n} \left( \langle \widehat{p_1}, \widehat{p_2}, \widehat{m_1}, \widehat{m_2} \rangle - \langle p_1,p_2, m_1, m_2 \rangle \right) \rightarrow N(0, \Sigma^2).
\end{align*}

To derive an approximation of the variance of $D(\widehat{m}_1, \widehat{m}_2)$, we let the $\vecdiv$ operator denote the division for every group $g$ of $m_g$ by $p_g$ to get $\mu_g$.
We can first apply the delta method just to the $\vecdiv$ operator, obtaining a simple asymptotic covariance matrix for the sample conditional means. 
The derivative of the $\vecdiv$ operator is made of two diagonal $2 \times 2$ blocks:
\begin{align*}
  \begin{bmatrix} P & M \\
  \end{bmatrix} &= 
  \begin{bmatrix}
  -m_1/p_1^2 & 0 & 1/p_1 & 0 \\
  0 & - m_2/p_2^2 & 0 & 1/p_2
  \end{bmatrix}
 \end{align*}
Hence the asymptotic covariance matrix obtained by the delta method is 
\begin{align*}
      \Lambda^2 &= \frac{1}{n} \begin{bmatrix} P & M \end{bmatrix} \begin{bmatrix} A & B \\ 
      B^T & C \\ 
      \end{bmatrix} \begin{bmatrix} P \\
      M \\
      \end{bmatrix} \\
\intertext{Some algebra (start by substituting $\mu_g$ for $m_g / p_g$) gets us the desired asymptotic covariance matrix for the sample conditional means.}
  \Lambda^2 &= \frac{1}{n} \begin{bmatrix} \frac{\sigma_1^2}{p_1} & 0 \\
  0   & \frac{\sigma_2^2}{p_2} \\
  \end{bmatrix}.
\end{align*}
Let $u = \sum_{j=1}^2 \mu_j$ and $v = \sum_{j=1}^2 \mu_j^2$.  
The partial derivative of $D(\mu_1, \mu_2)$ with respect to $\mu_g$, $g \in \{ 1, 2 \}$, is:
\begin{align}
  \frac{\partial}{\partial \mu_g} D(\mu_1, \mu_2) &= 4 \sum_{i=1}^2 \left( \frac{\mu_i}{\sum_{j=1}^2 \mu_j} - \frac{1}{2} \right) \left( \frac{\delta_{i g}}{\sum_{j=1}^2 \mu_j} - \frac{\mu_i}{(\sum_{j=1}^2 \mu_j)^2} \right) \notag \\
            &= 4 \frac{\mu_g - v/u}{u^2}.
\end{align}

Using the delta method for the composition of $D$ with the division operator can be done by multiplying the asymptotic covariance matrix of the conditional means, $\Lambda$, on its left and right by the derivative of $D$ and its transpose, effectively applying the chain rule.  We get the following large $n$ approximation $\sigma^2$ for the variance of $D(\widehat{\mu_1}, \widehat{\mu_2})$,
\begin{align*}
\sigma^2 &= \frac{16}{n u^4} 
\left< \mu_1 - \frac{v}{u}, \mu_2 - \frac{v}{u} \right>
 \begin{bmatrix} \frac{\sigma_1^2}{p_1} & 0 \\
  0   & \frac{\sigma_2^2}{p_2} \\
  \end{bmatrix}
    \left< \mu_1 - \frac{v}{u}, \mu_2 - \frac{v}{u} \right>^T\\
 &= \frac{16 (\mu_1 - \mu_2)^2}{n (\mu_1 + \mu_2)^6} \left( \mu_2^2 \left( \frac{\sigma_1^2}{p_1} \right) + \mu_1^2 \left( \frac{\sigma_2^2}{p_2} \right) \right) 
\end{align*}
\end{proof}

\section{Code Appendix}\label{app-code}

\begin{lstlisting}
##########################################
## Calculate DER statistic and variance ##
##########################################

library(data.table)

# Set constants (dummy values included here)
n_min = 0; n_max = 1
meanlog_min = 0; meanlog_max = 1
trt_effect_min = 0; trt_effect_max = 1
group_effect_min = 0; group_effect_max = 1
sdlog_min = 0; sdlog_max = 1
mixture_prob_min = 0; mixture_prob_max = 1

#' Calculate DER diff square measure from vector of means
#'
#' @param means vector of means for groups (length 2, each group once)
#' @return DER diff square measure
calculate_der_diff_vector <- function(means) {
  sum_means <- sum(means)
  der_diff <- 2 *  sum((means / sum_means - 1 / 2)^2)
  return(der_diff)
}

#' Calculate DER diff square measure from data.table of individual data
#' 
#' @param dt data.table, with columns "group" (2 groups) and "Y" (numerical data for the DER measure)
#' @return DER diff square measure
calculate_der_diff_indiv_datatable <- function(dt) {
  return(calculate_der_diff_vector(dt[, .(mean = mean(Y)), keyby = .(group)]$mean))
}

#' Calculate treatment - control difference of
#' DER diff square measures from data.table of individual data.
#'
#' @param dt data.table, with columns "variant", "group" (2 groups) and "Y" (numerical data for the DER measure)
#' @return diff of DER diff square
calculate_der_diff_diff_indiv_datatable <- function(dt){
  derdiff_c <- calculate_der_diff_indiv_datatable(dt[variant == "control",])
  derdiff_t <- calculate_der_diff_indiv_datatable(dt[variant == "treatment",])
  der_diff_diff <- derdiff_t - derdiff_c
  return(der_diff_diff)
}

#' Delta method variance estimate of DER diff square measure from vectors with group size, mean, and variance, groups labeled by indices; length 2, one index per group.
#'
#' @param group_ns vector of number of members of each group
#' @param means vector of means of groups
#' @param variances vector of variances of groups
#' @return delta method estimate of variance
delta_estimate_der_diff_variance_vectors <- function(group_ns, means, variances){
  n <- sum(group_ns)
  u <- sum(means)
  v <- sum(means^2)
  w <- means[1]
  x <- means[2]
  
  # group fractions
  p <- group_ns / n
  
  val <-  (16 * (w - x)**2) / (n * (w + x)**6)
  sum_vars <- (w**2 * variances[2]/p[2]) + (x**2 * variances[1]/p[1])
  
  variance_est <- val * sum_vars
  
  return(variance_est)
}

#' Delta method variance estimate of DER diff square measure and its variance from data.table of individual data.
#' 
#' @param dt data.table, with columns "group" and "Y" (numerical data for the DER measure)
#' @return list with DER: mean of DER and var: variance of DER for that data table
delta_estimate_der_diff_variance_indiv_datatable <- function(dt){
  dt_agg <- dt[, .(N = .N, mean = mean(Y), var = var(Y)), keyby = .(group)]
  return(delta_estimate_der_diff_variance_vectors(dt_agg[["N"]], dt_agg[["mean"]], dt_agg[["var"]]))
}

#' Calculate delta method variance estimate for treatment - control diff of DER diff square measure, from a data.table with individual member metric data.
#'
#' @param dt data.table, with columns "variant", "group" (no groups omitted)
#'           and "Y" (numerical data for the DER measure)
#' @return variance estimate
delta_estimate_der_diff_diff_variance_indiv_datatable <- function(dt){
  var_c <- delta_estimate_der_diff_variance_indiv_datatable(dt[variant == "control",])
  var_t <- delta_estimate_der_diff_variance_indiv_datatable(dt[variant == "treatment",])
  var_diff <- var_t + var_c
  return(var_diff)
}

#' Calculate treatment - control diff of DER diff square measure and variance estimate, from a data.table with individual member metric data.
#'
#' @param dt data.table, with columns "variant", "group" (2 groups),
#'           and "Y" (numerical data for the DER measure)
#' @return list of DER statistic and variance estimate
der_results <- function(dt){
  der_statistic <- calculate_der_diff_diff_indiv_datatable(dt)
  variance_estimate <- delta_estimate_der_diff_diff_variance_indiv_datatable(dt)

  der_result_list <- list("der_statistic" = der_statistic,
                          "variance_estimate" = variance_estimate)
}

#' Function to generate data with a given sample size, mean, standard deviation, and mixture probability
#'
#' @param sample_size total number of samples to draw. Defaults to NULL.
#' @param meanlog mu parameter for log-Normal, can be any real number. Defaults to NULL.
#' @param trt_effect treatment effect, can be any real number. Defaults to NULL.
#' @param group_effect difference in metric mean by group. Defaults to NULL.
#' @param sdlog sigma parameter for log-Normal, can be any positive real number. Defaults to NULL.
#' @param mixture_prob mixture probability, taking values strictly between 0 and 1. Defaults to NULL.
#' @return Vector of mock LinkedIn data following a log-Normal distribution mixed with a point mass at 0, drawn with the specified parameters, and rounded to the nearest integer.
generate_li_data <- function(sample_size = NULL, meanlog = NULL, trt_effect = NULL, group_effect = NULL, sdlog = NULL, mixture_prob = NULL){
  # draw all parameters that were not specified
  sample_size <- ifelse(is.null(sample_size), round(runif(1, min = n_min, max = n_max)), sample_size)
  meanlog <- ifelse(is.null(meanlog),  runif(1, min = meanlog_min, max = meanlog_max), meanlog)
  trt_effect <- ifelse(is.null(trt_effect), runif(1, min = trt_effect_min, max = trt_effect_max), trt_effect)
  group_effect <- ifelse(is.null(group_effect), runif(1, min = group_effect_min,, max = group_effect_min,), group_effect)
  sdlog <- ifelse(is.null(sdlog), runif(1, min = sdlog_min, max = sdlog_max), sdlog)
  mixture_prob <- ifelse(is.null(mixture_prob), runif(1, min = mixture_prob_min, max = mixture_prob_max), mixture_prob)
  
  # draw variant and group randomly
  variant <- sample(c("treatment", "control"), size = sample_size, replace = TRUE)
  group <- rbinom(sample_size, 1, 0.5)
  
  # define vector of individual means and draw shared standard deviation
  # note that treatment and group effects are scaled to match log-normal expectations.
  full_mean <- meanlog + log(trt_effect + 1) * (variant == 1) + log(group_effect + 1) * (group == 1)
  
  Y <- round(rlnorm(sample_size, meanlog = full_mean, sdlog = sdlog) * rbinom(sample_size, 1, mixture_prob))
  
  dt <- data.table("Y" = Y,
                   "variant" = variant,
                   "group" = group)
  
  parameters <- list("sample_size" = sample_size,
                     "meanlog" = meanlog,
                     "trt_effect" = trt_effect,
                     "group_effect" = group_effect,
                     "sdlog" = sdlog,
                     "mixture_prob" = mixture_prob)
  
  return_value <- list("dt" = dt,
                       "parameters" = parameters)
  
  return(return_value)
}

# Example usage
li_data <- generate_li_data()
dt <- li_data$dt
der <- der_results(dt)
\end{lstlisting}

\end{document}